\documentclass[journal,twoside]{IEEEtran}
\usepackage{graphicx}
\DeclareGraphicsExtensions{.pdf}
\usepackage{caption}
\usepackage{subcaption}
\newcommand{\bea}{\begin{eqnarray}}

\newcommand{\eea}{\end{eqnarray}}

\newcommand{\rmd}{{\rm d}}

\newcommand{\Beta}{{\rm B}}

\begin{document}

\markboth{L. R. Arnaut -- 10 July 2014}{L. R. Arnaut -- 10 July 2014}


\title{Design of Chaotic Cavities with Curved Wave Diffractors for Enhanced Low-Frequency Operation
}

\author{
	L. R. Arnaut\thanks{L. R. Arnaut is with the Institute of Electronics and Telecommunications Research (IETR), UMR CNRS 6164, National Institute of Applied Sciences (INSA), F-35708 Rennes Cedex 7, France.}
}

\maketitle

\date{10 July 2014}

\begin{abstract}
Some numerical calculations are presented on the dependence of the average mode count and average mode density of electromagnetic cavities on their specific geometric design, based on the generalized Weyl law. The analysis focuses on a chaotic quasi-cubic cavity furnished with curved wave diffractors placed on its interior surface. The focus is on a design that increases the mode density and $Q$ at relatively low frequencies. The results are of interest in reducing the `lowest usable frequency' and increasing the maximum field strength inside mode-stirred reverberation chambers.

\end{abstract}

\section{Introduction \label{sec:intro}}
Intrinsic chaoticity of a static cavity can be achieved by applying geometrical modifications to a `regular' (i.e., integrable) rectangular cavity that involve adding curved surfaces, provided the mean curvature is strictly negative. Classical examples in two dimensions are the Sina\"{i} billiard and the Bunimovich stadium. 
The resulting change in interior shape affects the mode density across the entire cavity spectrum, in particular at intermediate to low frequencies (nonasymptotic regime of the generalized Weyl density \cite{weyl1}--\cite{luko1}). On account of the spectral probability distributions of widths of isolated resonances \cite{port1}, their  spectral spacings \cite{bohi1}, and their ratio (quantifying the modal overlap \cite{cela1}), this change affects the modal coupling. As a result, this influences the distribution and statistics of the quality factor $Q$ of the cavity, because its PDF $f_Q(q)$ contains a degrees-of-freedom parameter that depends on the number of simultaneously excited modes \cite{arnaTEMC_Q}. 

The question regarding the effect of shape on $Q$ can be approached in two ways:
\begin{enumerate}
\item {\it numerically}: using full-wave simulation, one calculates and spatially integrates the interior fields (for stored energy) and the induced currents in the cavity walls (for dissipated power), then repeats this for each stir state, from which the random stored and dissipated energies follow, yielding the PDF and statistics of $Q$;
\item {\it analytically}: using the expressions for the known average mode density and its dependence on dimensions, shape parameters (including radii of curvature, edge length, etc.), one estimates the effect of one or multiple spherical caps on the distribution parameter $M$ in $f_Q(q)$ and on the mean, standard deviation and confidence interval for $Q$ in particular.
\end{enumerate}

The numerical simulation of stirred electrically large 3-D cavities is very time consuming. Also, it suffers from inaccuracies and artefacts on the mesh introduced by staircasing effects when approximating curvature of diffracting surfaces. Therefore, in order to get a first estimate but which includes the quantified effect of the curved surfaces on the statistics of $Q$, we shall perform the latter approach first.

This report builds on an earlier analysis and discussion \cite{arnaTEMC_LUF}.

\section{Effect of hemispherical surfaces on low-frequency modal density}
\subsection{Average mode count of parallelepiped cavity with hemispherical diffractors}
The generalized Weyl law reads \cite{weyl1}--\cite{luko1} (cf. \cite{arnaTEMC_LUF} for definition of symbols)
\bea
N(f) &=&  \frac{8\pi V}{3 c^3} f^3 - \left [ \frac{4}{3\pi c} \int\int_{S=\partial V} \frac{\rmd s}{\varrho(s)} 
\right. \nonumber\\  &~& \left.
- \frac{1}{6\pi c} \int_{L} \frac{[\pi - \phi(\ell)] [\pi - 5 \phi(\ell)]}{\phi(\ell)} \rmd \ell\right ] f.
\label{eq:Weylgen}
\eea
For convex (`inward bulging') type curved diffractors, when viewed from inside the cavity, the radius of curvature $\varrho$ is defined as being negative \cite{luko1}.

Ignoring\footnote{For hemispheres, the additional dihedral angles are $\pi/2$, which give rise to a reduction in mode density. This can be avoided by `filling in' these edges, thus making the transitions between walls and diffractor surface smooth.} for the moment the effect of diffractors on the edge term (line integral in (\ref{eq:Weylgen})), we focus on
the effect of this curvature, it is observed that \cite{arnaTEMC_LUF}:
\begin{itemize}
\item a negative value of the curvature $1/\varrho$ for convex diffractors produces an increase of mode density. The smaller the modulus of the local radius of curvature, $|\varrho|$, the larger its effect on the mode density ${\rm d}N(f)/{\rm d}f$;
\item the excision of the interior volume by adding convex diffractors ultimately reduces the mode density when frequency is sufficiently high (so that the leading volume term dominates). However, this occurs at the benefit of increased mode density at relatively low frequencies;
\item the effect of curvature and volume reduction on mode count and mode density are proportional to the number of diffractors, $n$, and their spatial density. For a footprint area $\pi |r|^2$ of a single diffractor, it follows that asymptotically $n \propto 1/|r|^2$ and ${\rm d}N(f)/{\rm d}f \propto 1/|r|$.
\end{itemize}
Thus, an array of many small but strongly curved surface diffractors (`egg crate'-style) has a stronger effect on mode density than one large diffractor because of all three reasons.  

Physically, the increase in low-frequency mode count introduced by curvature may be  
understood by the fact that such additional curvature generates more complex spatial current distributions on the interior surface, compared to those on a flat surface (cf. modal current distribution in a rectangular cavity). These current patterns are related to the interior field distribution, on account of the boundary conditions. 

For simplicity of calculation, we analytically investigate hemispherical caps of height $h=|R|$ rising above the original flat cavity surface. Extension to more general values of $h$ is done through numerical results. The radius $|R|\equiv |\varrho|$ is chosen as a function of the intended frequency range of operation (noting that, for fixed $n$, lower frequencies require larger $|R|$ in order to be more effective), the available cavity volume for placing diffractors (which reduces the working volume), the density of wall or volume diffractors, etc.

For a single hemisphere on a flat cavity wall, the reduction of the surface area of the cavity wall by orthogonal projection of the diffractor's surface onto this wall (area $\pi R^2$) is overcompensated by the area of the hemispherical surface itself (area $4\pi R^2/2$), yielding a net increase $\Delta S = \pi R^2$. The cavity volume decreases by the volume enclosed by the hemisphere and the wall, i.e., $(4\pi R^3/3)/2$. The increase of edge length per hemisphere is that of its rim, i.e., $2\pi |R|$, with associated dihedral angle $\pi/2$. The radius of curvature is constant across the hemisphere, i.e., $\varrho = -|R|$. 

For $n$ identical hemispherical wall diffractors, (\ref{eq:Weylgen}) becomes (cf. Appendix, eq. (\ref{eq:modecountsinglediff}))
\bea
N(f) &=& 
\frac{8\pi}{3} \left ( \frac{f}{c} \right )^3 \left ( V - n \frac{2 \pi |R|^3}{3}   \right ) 
 \nonumber\\  &~&
- \left [ L+W+H-n\frac{(16-3\pi) |R|}{6} \right ] \frac{f}{c} .
\label{eq:Weylgentwo}
\eea
The absolute change in the average mode count as a result of adding convex diffractors, $\Delta N \stackrel{\Delta}{=} N_{\rm chao}-N_{\rm reg}$, given by 
\bea
\Delta N = - n \left [ \frac{16\pi^2 |R|^3}{9} \left ( \frac{f}{c} \right )^3 - \frac{(16-3\pi)|R|}{6} \left ( \frac{f}{c} \right ) \right ].\label{eq:Weyldiffracfinal_copy}
\eea
may become positive at sufficiently low frequencies.
In (\ref{eq:Weylgentwo}) and (\ref{eq:Weyldiffracfinal_copy}), the terms `$3\pi$' are due to the dihedral right angle between the diffractor's and the cavity's surfaces ($\phi = \pi/2$). For this value of $\phi$, these reduce the mode density. These terms can be made to vanish -- thus increasing the LF mode density -- by rounding this angle, i.e., $\phi = \pi$. A further increase in the low-frequency mode density can be achieved by similarly rounding the eight original edges of the rectangular cavity, along $L$, $W$ and/or $H$.
Note that this introduces concave curvature and, hence, produces a focusing effect at high frequencies. The radius of curvature for these roundings should therefore be kept low to avoid this.  

From (\ref{eq:Weyldiffracfinal}), it follows that if
\bea
0 < \frac{f |R|}{ c} < \sqrt{\frac{3(16-3\pi)}{32\pi^2}} \simeq 0.2499
\eea
i.e.,
\bea
|R| < \lambda/4
\eea
then an increase of the average mode density may arise, depending on the sign of the denominator in (\ref{eq:relativechangemodedenshemi}), viz. provided the wavelength is sufficiently small. In practice, this is always satisfied for the cavity under investigation (cf. Appendix).

\section{Some Calculations on Parametric Dependence for the Medium Sized Cavity}
\subsection{Empty Cavity}
\begin{figure}[htb] \centering \begin{tabular}{c}
\vspace{-4cm}\\
\hspace{-1.2cm}
\includegraphics[scale=0.5]{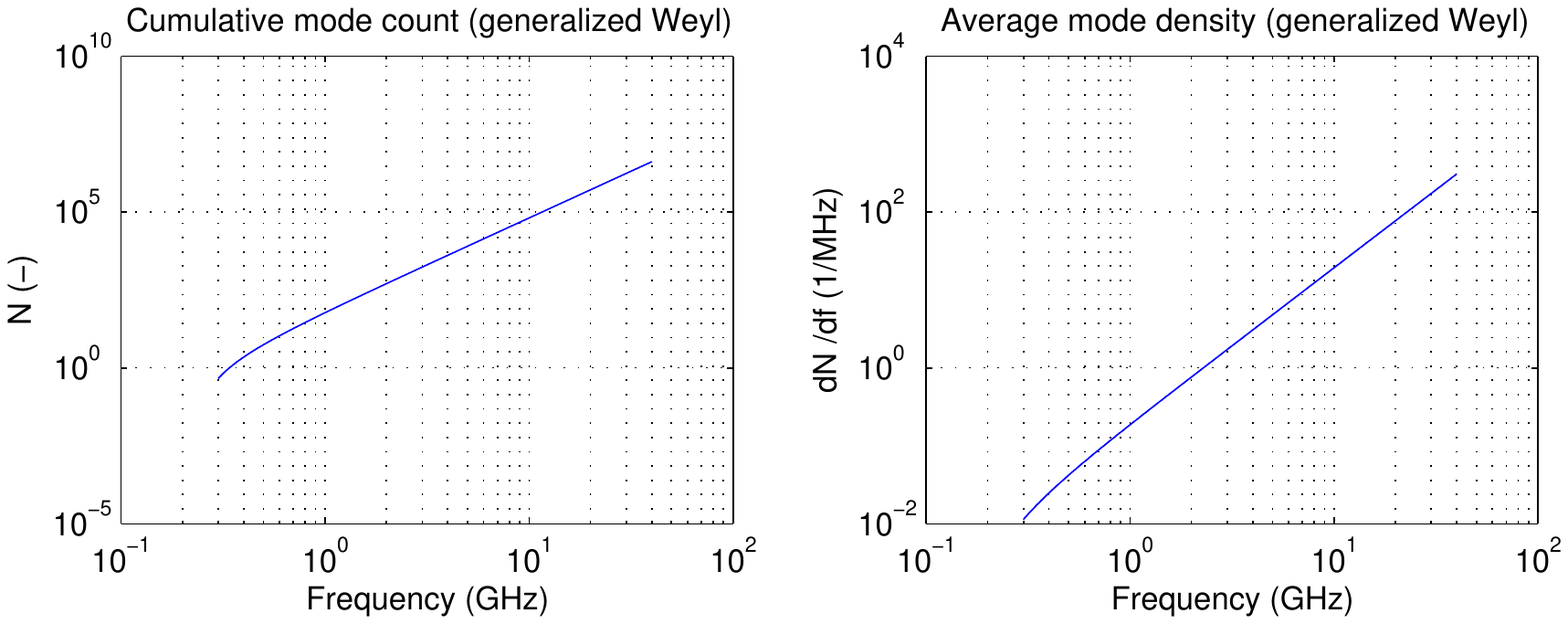}\ \\
\vspace{-7.5cm}\\
\hspace{-1.2cm}
(a)\\
\\
\vspace{-4cm}\\
\hspace{-1.2cm}
\includegraphics[scale=0.5]{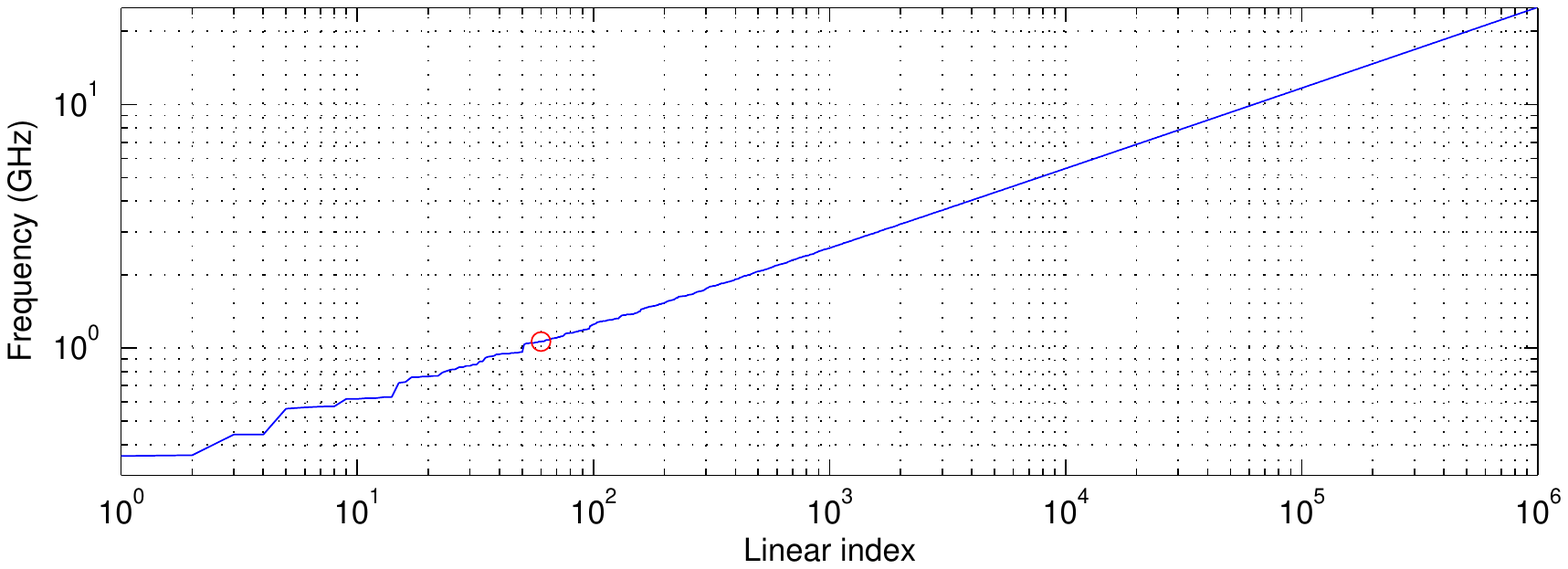}\ \\
\vspace{-7.5cm}\\
\hspace{-1.2cm}
(b)\\
\vspace{0cm}\\
\\
\end{tabular}
{
\caption{\label{fig:MED_resfreqs_0to25GHz} \small
(a) Average mode count $N(f)$, average mode density $\rmd N(f) / \rmd f$ from $0$ to $40$ GHz and
(b) actual cumulative mode count (for linearized index number of $10^6$ mode triplets $mnp$) for nonchaotic lossless rectangular cavity of dimensions $L=60$ cm, $W=59$ cm, $H=58$ cm from $0$ to $25$ GHz. The red circle indicates the location of the $60$th mode.}
}
\end{figure}

Consider a cavity of dimensions $L\times W \times H=60\times 59\times 58$ cm$^3$.
Fig. \ref{fig:MED_resfreqs_0to25GHz} shows the cumulative mode count, mode density up to $40$ GHz and resonance frequencies up to $25$ GHz. Also marked is the $60$th mode at $1.1211$ GHz, which provides an order-of-magnitude estimate for the LUF of an electronically stirred cavity.

\begin{figure}[htb] \begin{center} \begin{tabular}{c}
\vspace{-4cm}\\
\hspace{-1.2cm}
\includegraphics[scale=0.5]{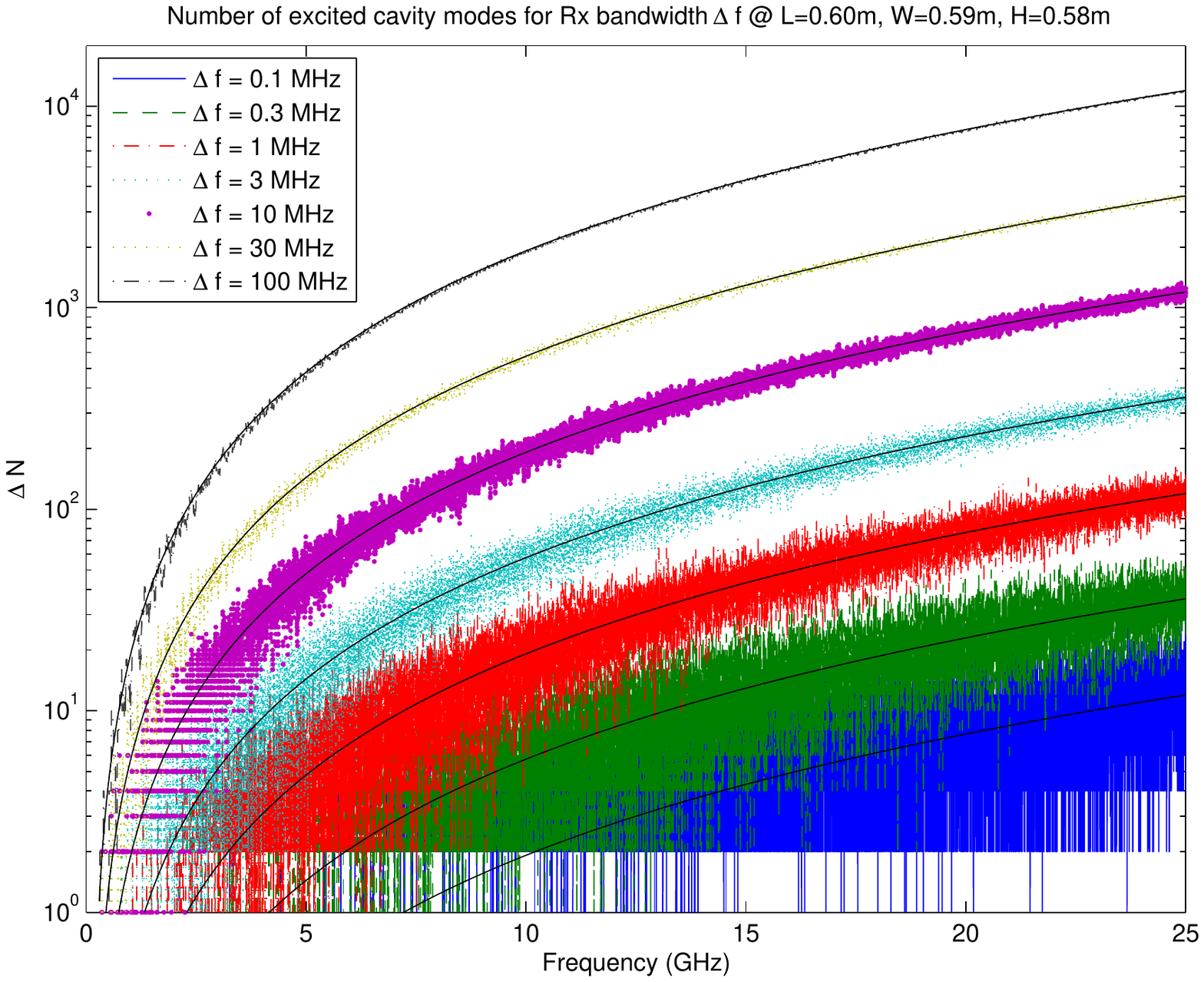}\ \\
\vspace{-4cm}\\
\hspace{-1.2cm}
(a)\\
\\
\vspace{-4cm}\\
\hspace{-1.2cm}
\includegraphics[scale=0.5]{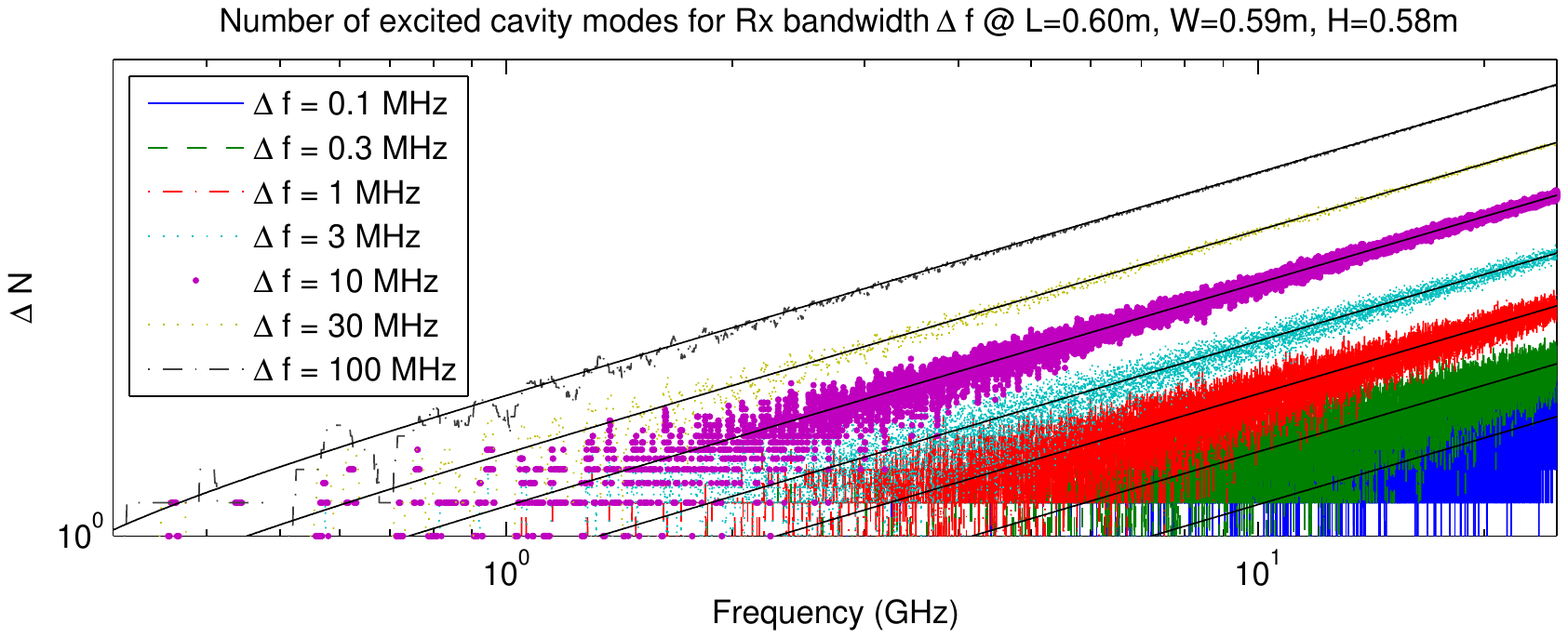}\ \\
\vspace{-7.5cm}\\
\hspace{-1.2cm}
(b)\\
\\
\vspace{-4cm}\\
\hspace{-1.2cm}
\includegraphics[scale=0.5]{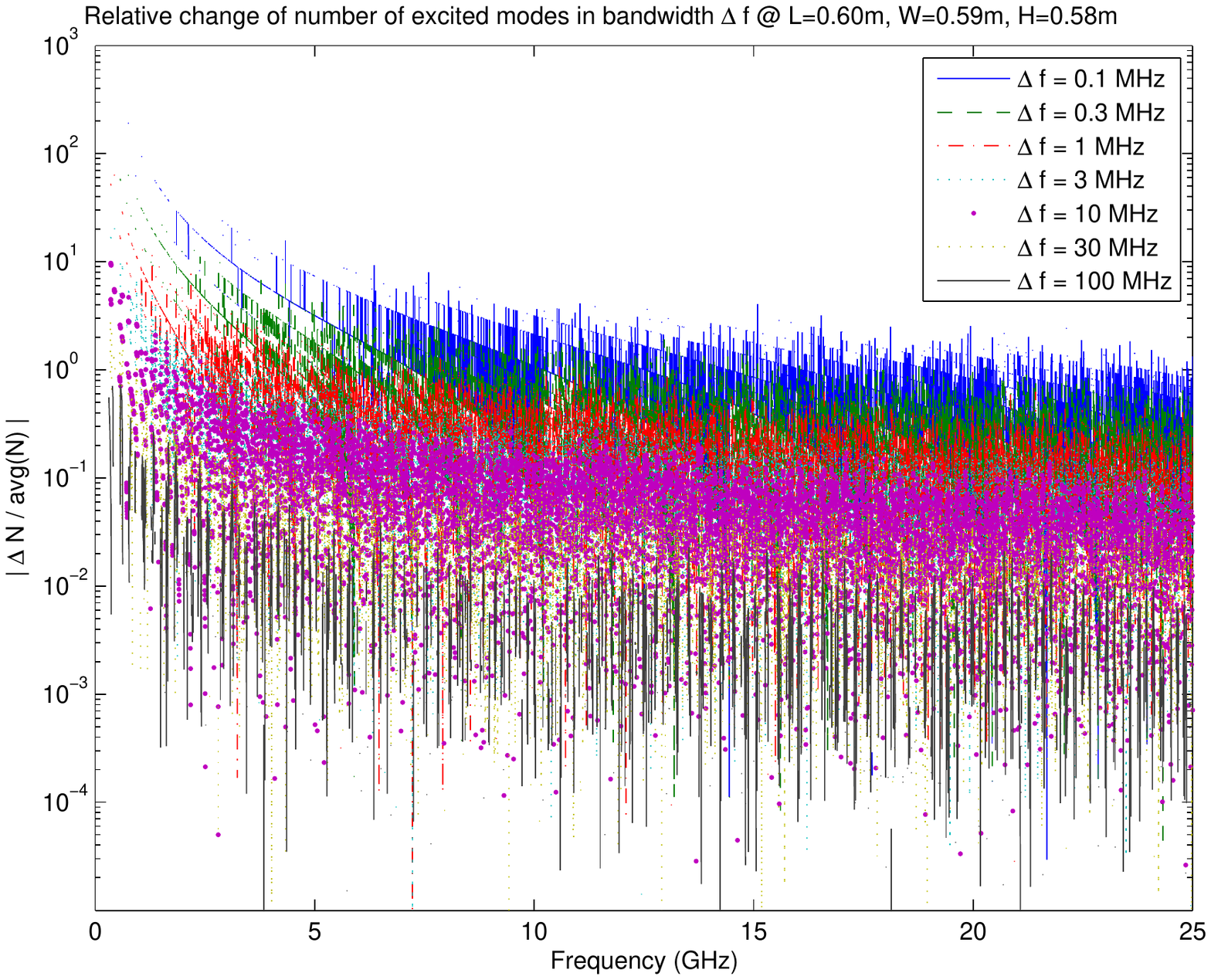}\ \\
\vspace{-4cm}\\
\hspace{-1.2cm}
(c)
\end{tabular}
\end{center}
{
\caption{\label{fig:participatingmodes_relchange_0to25GHz} \small
(a) Number of excited modes $\Delta N(f)$ for nonchaotic lossless rectangular cavity of dimensions $L=60$ cm, $W=59$ cm, $H=58$ cm, at selected values of receiver bandwidth $\Delta f$. 
Colored lines: based on calculated actual spectrum of cavity modes, $N(f+\Delta f/2)-N(f-\Delta f/2)$. Black lines: estimated values based on generalized Weyl law (smoothed average mode density), $(\rmd N(f)/\rmd f)\Delta f$. 
(b) Same as (a) on logarithmic scale.
(c) Relative fluctuations of number of excited modes $\Delta N(f)$ around Weyl average: $|\Delta N | / \langle N \rangle$.}
}
\end{figure}

Fig. \ref{fig:participatingmodes_relchange_0to25GHz} show that the actual number of contributing modes strongly depends on the effective cavity bandwidth (due to modal overlap) or the receiver bandwidth, whichever is the greater.
It also shows that the relative fluctuations of the actual mode density around the asymptotic law remain substantial even up to high frequencies. However, in the case of mod-stirred reverberation chambers, it can be expected that these fluctuations when considered across an ensemble of stir states are much reduced through ensemble averaging. This explains why we focus on this average density, rather than on the actual density or its spectral fluctuations.

\begin{figure}[htb] \begin{center} \begin{tabular}{c}
\vspace{-4cm}\\
\hspace{-1.2cm}
\includegraphics[scale=0.5]{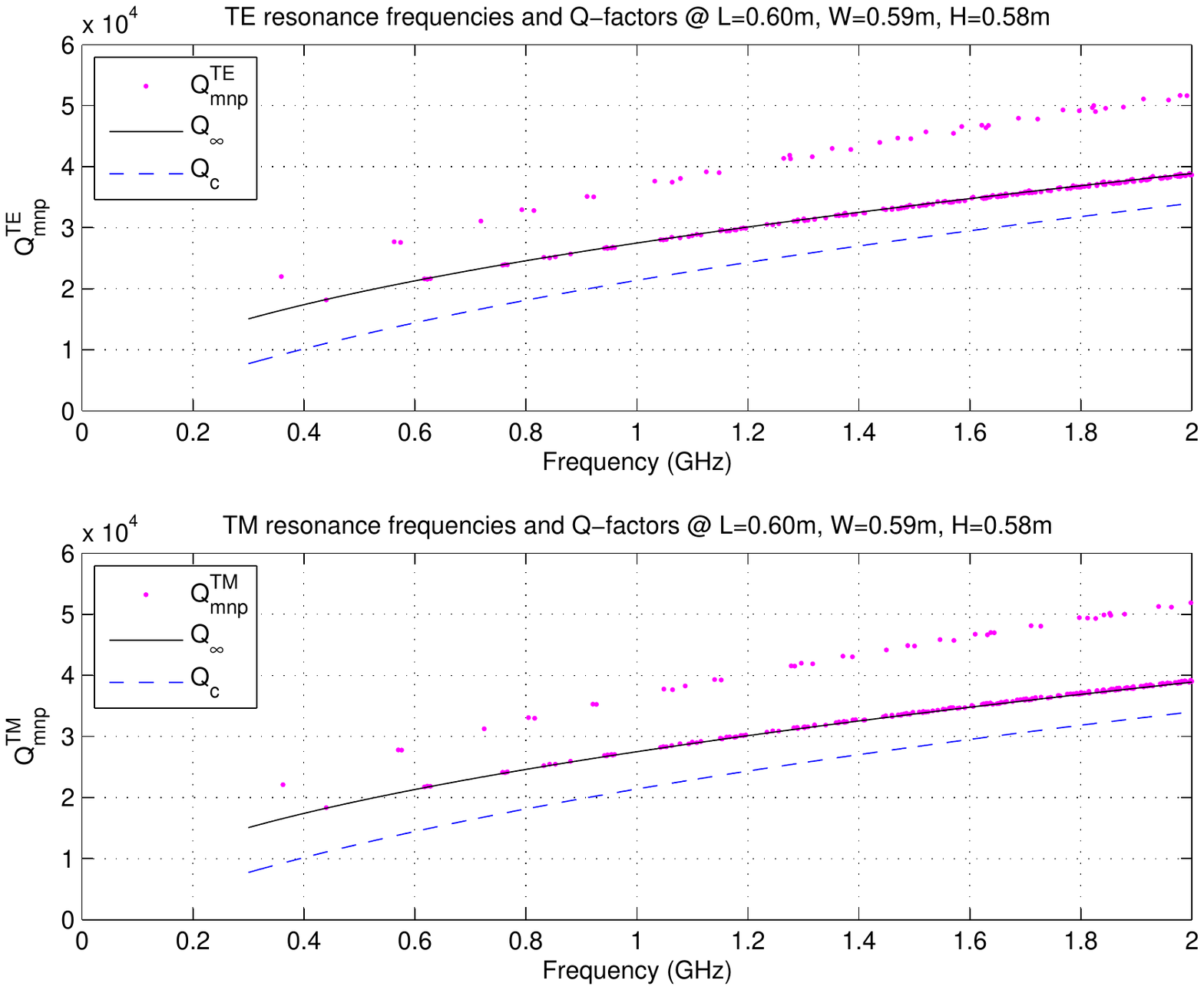}\ \\
\vspace{-4cm}\\
\hspace{-1.2cm}
(a)\\
\\
\vspace{-4cm}\\
\hspace{-1.2cm}
\includegraphics[scale=0.5]{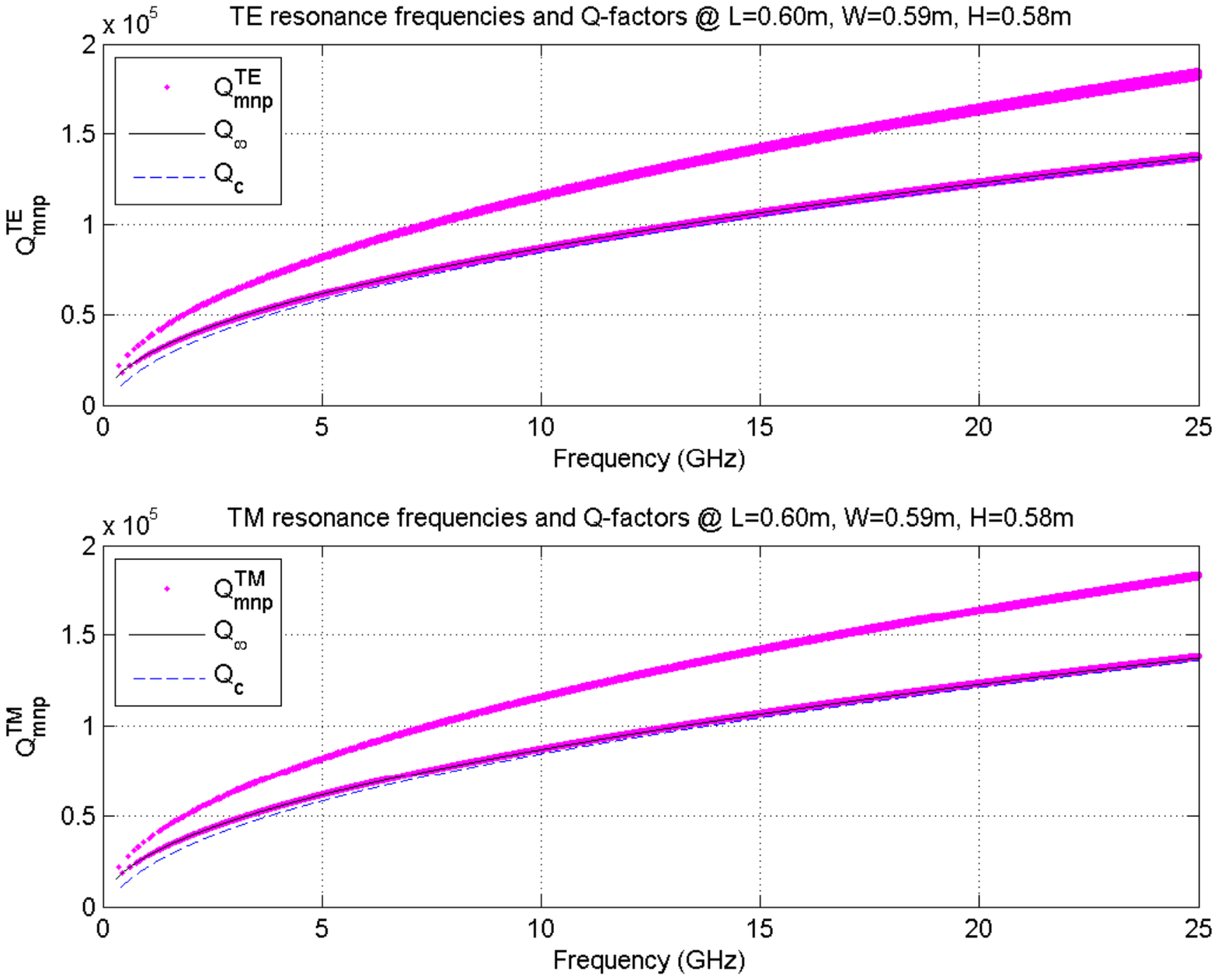}\ \\
\vspace{-4cm}\\
\hspace{-1.2cm}
(b)\\
\\
\vspace{-4cm}\\
\hspace{-1.2cm}
\includegraphics[scale=0.5]{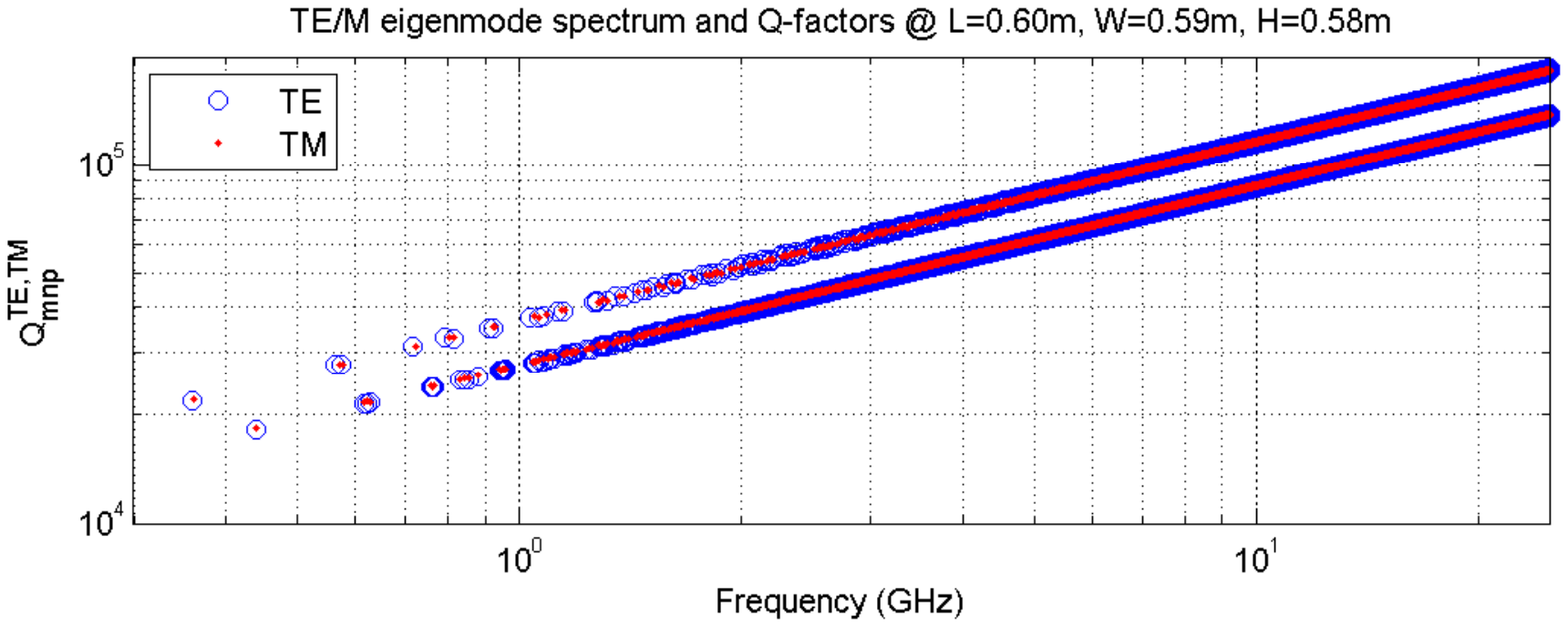}\ \\
\vspace{-7.5cm}\\
\hspace{-1.2cm}
(c) \\
\\
\vspace{-4cm}\\
\hspace{-1.2cm}
\includegraphics[scale=0.5]{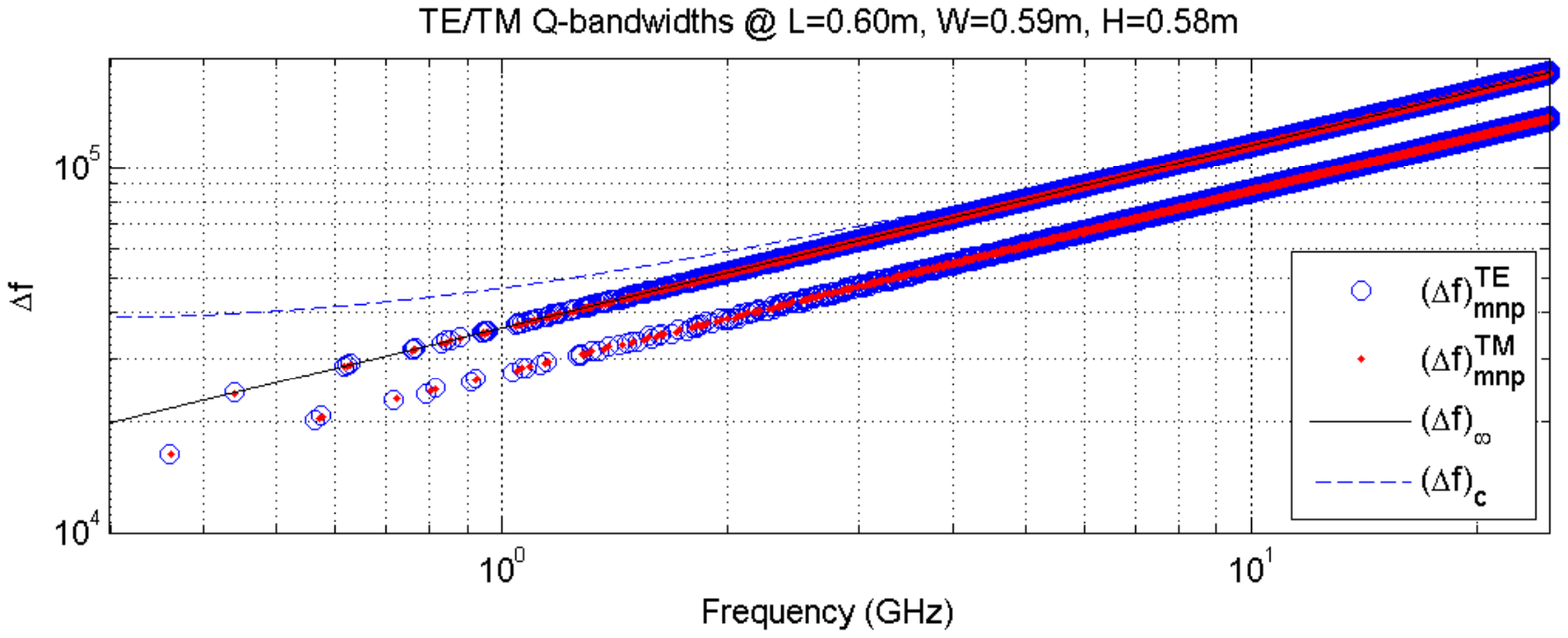}\ \\
\vspace{-7.5cm}\\
\hspace{-1.2cm}
(d)\\
\\
\vspace{-1cm}\\
\end{tabular}
\end{center}
{
\caption{\label{fig:TETMeigenmodespectrum_3MHz_3GHz} \small
TE and TM eigenmode spectra, modal quality factors, composite $Q_c$ and asymptotic $Q_\infty$ for aluminium rectangular cavity of dimensions $L=60$ cm, $W=59$ cm, $H=58$ cm: (a) $300$ MHz to $2$ GHz, (b)-(c) $300$ MHz to $25$ GHz, (d) associated effective $Q$ bandwidth for $f=300$ MHz to $25$ GHz.}
}
\end{figure}

Fig. \ref{fig:TETMeigenmodespectrum_3MHz_3GHz} shows TE/TM modal and effective ($Q_c$) quality factors between $0$ and $25$ GHz.
It is verified that the composite $Q_c$ does not interpolate between the degenerate and nondegenerate TE or TM modes. This is due to the fact that $Q_c$ assumes a continuous distribution of eigenvalues (resonances). The spectral gaps at low frequencies contribute with zero value of $Q$ in this spectral distribution, causing the $Q$ factor to be even smaller than the asymptotic $Q_\infty$.
By contrast, the average $\langle Q \rangle$ that follows from statistical analysis \cite{arnaTEMC_Q} produces values that are higher than $Q_\infty$ that would interpolate between degenerate and nondegenerate modes. 
At the same time, these values also correctly approach $Q_\infty$ when the weight by nondegenerate modes (no zero modal indices) dominates the vanishingly small relative contributions by degenerate modes when $f \rightarrow +\infty$.

\subsection{Cavity with Spherical Cap Diffractors}
\subsubsection{Effect of Changing $|R|$ or $h/|R|$ on $N(f)$ for Fixed Frequency}
\begin{figure}[htb] \begin{center} \begin{tabular}{c}
\vspace{-4cm}\\
\hspace{-1.2cm}
\includegraphics[scale=0.5]{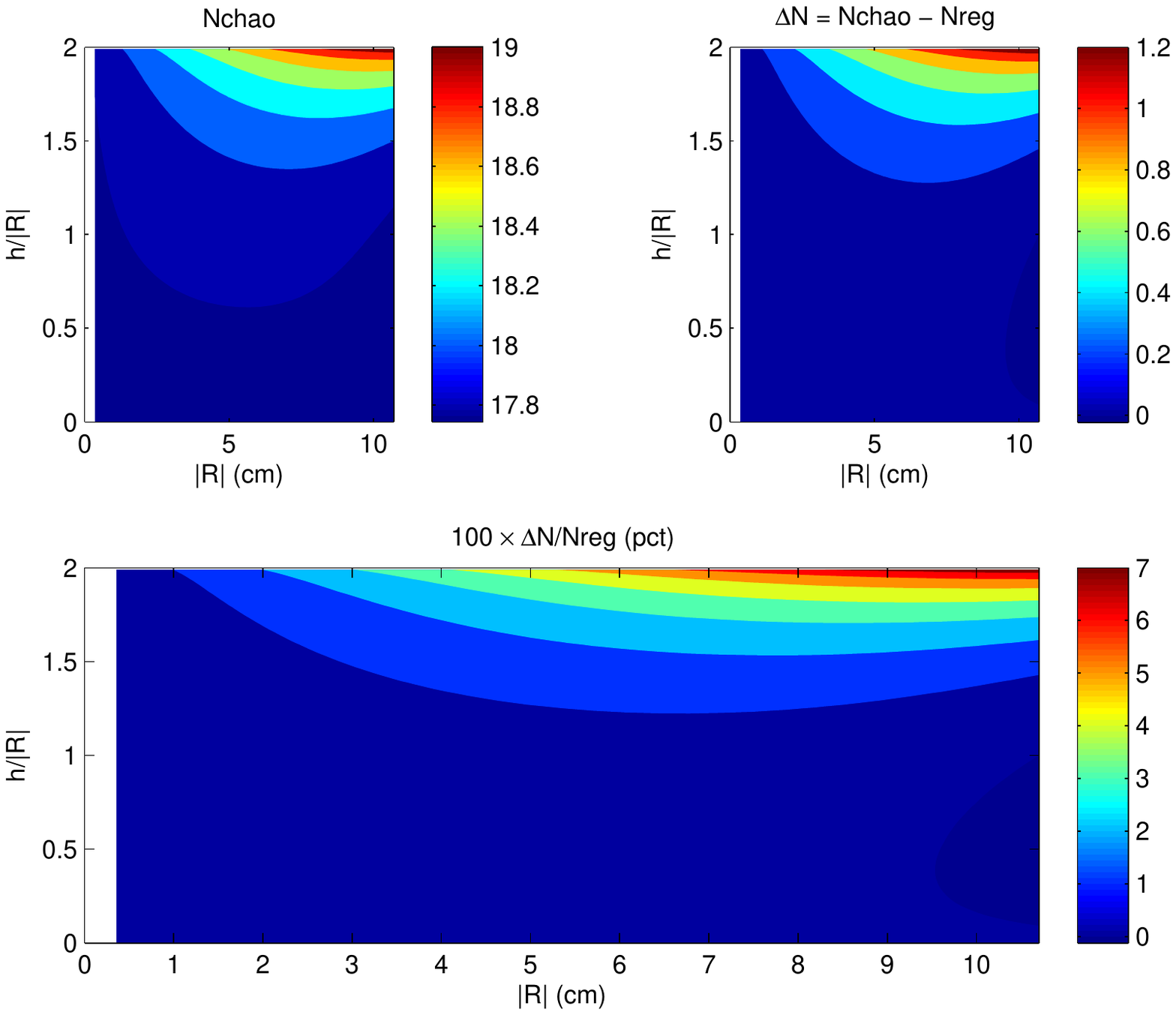}\
\vspace{-4cm}\\
\hspace{-1.2cm}
\end{tabular}
\end{center}
{
\caption{\label{fig:contourplot1} \small
Cumulative average mode count $N(f)$ at $f=700$ MHz for lossless rectangular cavity of dimensions $L=60$ cm, $W=59$ cm, $H=58$ cm furnished with single $(n=1)$ convex spherical cap diffractor having radius of curvature $-|R|$ and profile height $h$.}
}
\end{figure}

\begin{figure}[htb] \begin{center} \begin{tabular}{c}
\vspace{-4cm}\\
\hspace{-1.2cm}
\includegraphics[scale=0.5]{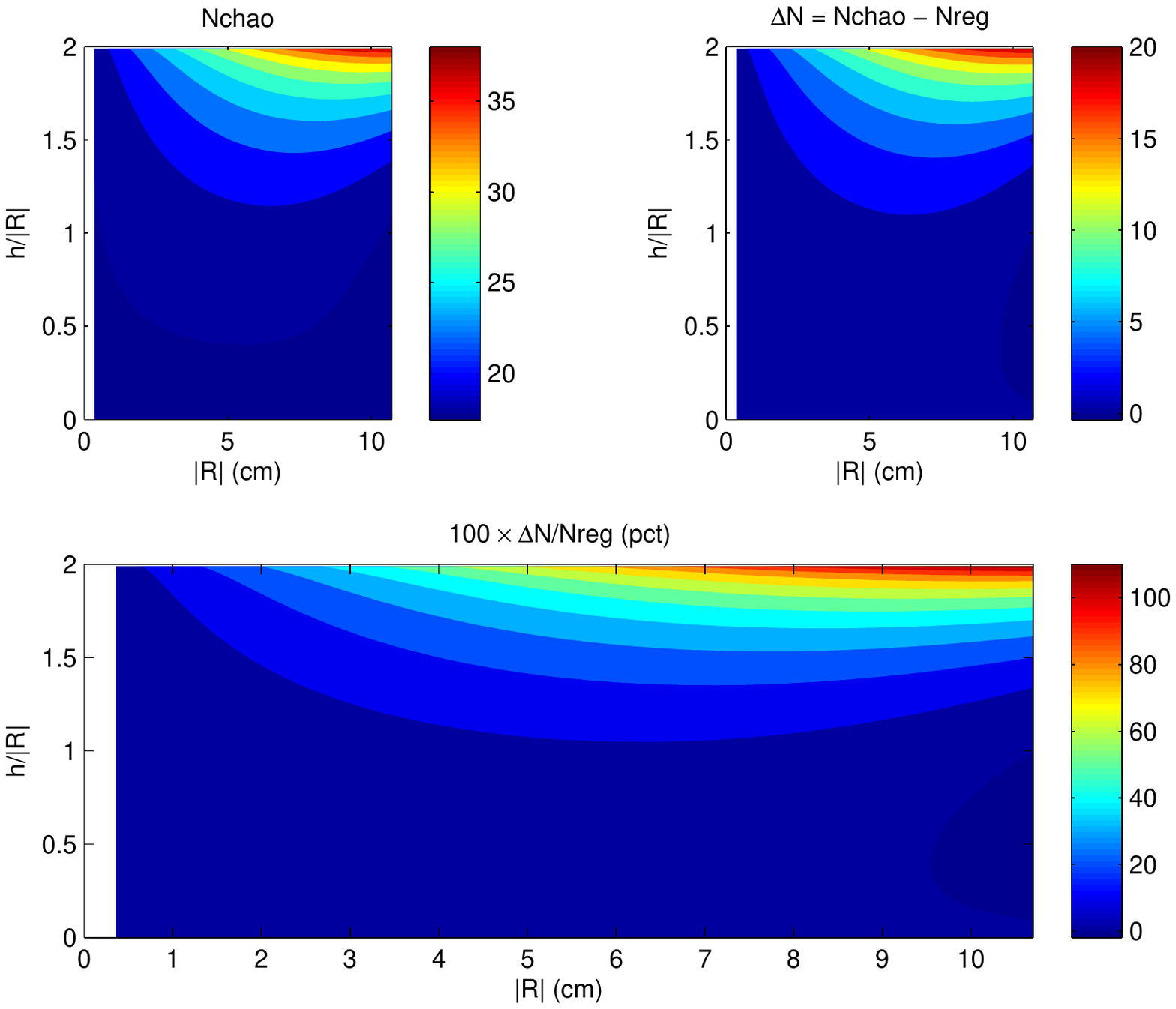}\ \\
\vspace{-4cm}\\
\hspace{-1.2cm}
\end{tabular}
\end{center}
{
\caption{\label{fig:contourplot15} \small
Same as Fig. \ref{fig:contourplot1} but for $n=15$ identical diffractors.}
}
\end{figure}

Fig. \ref{fig:contourplot1} shows the dependence of the cumulative mode count $N$ for empty (regular) cavity, chaotic cavity with a single diffractor ($n=1$), and the percentage increase of $N$, as a function of the radius of curvature $|R|$ and the normalized height of the diffractor $h/|R|$, 
where $h/|R|=1$ corresponds to a hemispherical diffractor. Since the $60$th mode of the empty cavity is at $1.12$ GHz, we seek to improve the mode density (well) below this anticipated LUF. 
To this end, the critical frequency for the diffractor design was set\footnote{Another choice would be $f_{\rm crit} = 1.12$ GHz.} as $f_{\rm crit}=700$ MHz, below (above) which an increase (decrease) of the average mode density arises as a result of adding diffractors. 
Note that this choice affects the maximum range of calculated values of $N$ for $|R|\leq 0.2499 ~\lambda_{\rm crit}$ and also affects the optimum value of $h/|R|$; see below.
From Fig. \ref{fig:contourplot1}, it can be seen that large ratios $h/|R|$ are needed for a single diffractor ($h/|R|\sim 2$, i.e., quasi-fully spherical `mushroom'-style) in order to achieve a noticeable increase of $N$.
However, if the number of diffractors is increased to $n=15$ (e.g., a $5 \times 3$ array of `large' diffractors, each with $|R|=5.75$ cm), the increase of $N$ is more substantial, as seen in Fig. \ref{fig:contourplot15} and {\it a fortiori} for three such arrays ($n=45$),
e.g., placed on mutually orthogonal walls.
Interestingly, the relative placement of these arrays (i.e., whether on same or orthogonal walls) is not relevant in this formulation.

\begin{figure}[htb] \begin{center} \begin{tabular}{c}
\vspace{-4cm}\\
\hspace{-1.2cm}
\includegraphics[scale=0.5]{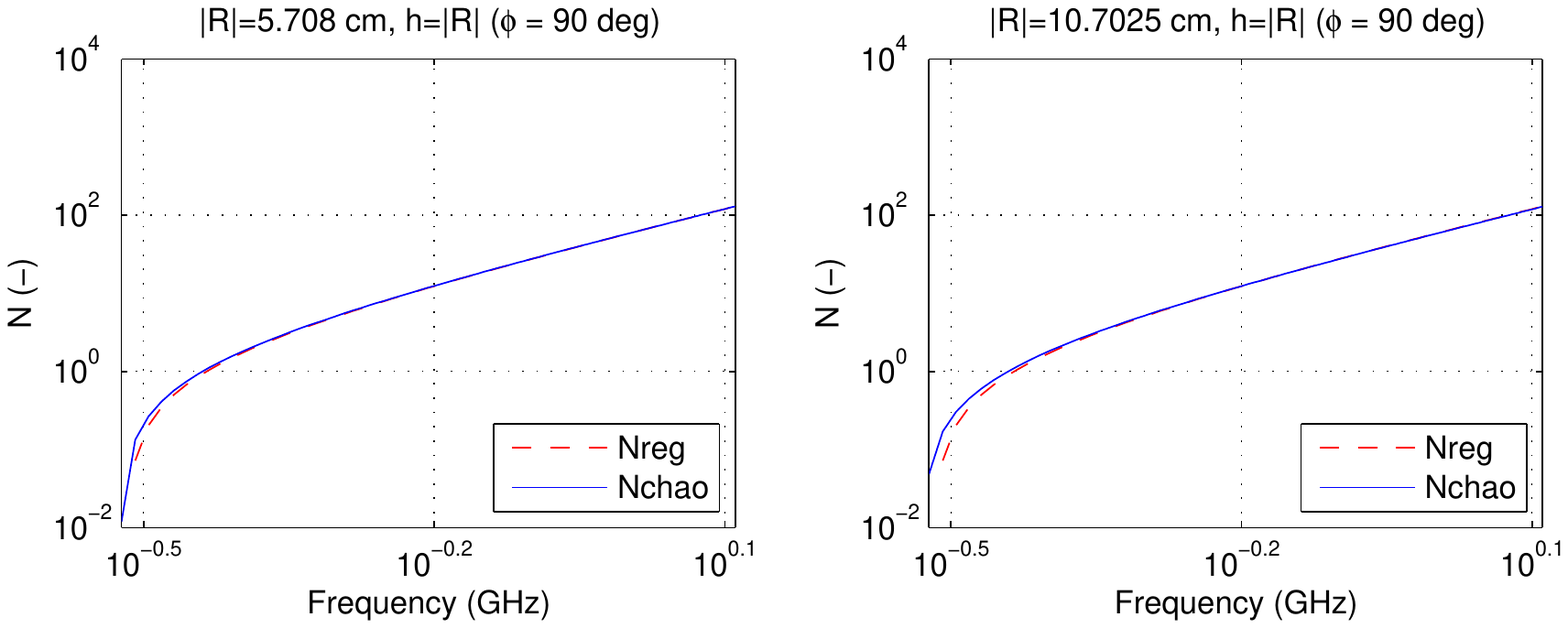}\
\vspace{-4cm}\\
\hspace{-1.2cm}
\end{tabular}
\end{center}
{
\caption{\label{fig:Nvsf1} \small
Frequency dependence of cumulative average mode count $N$ for lossless rectangular cavity of dimensions $L=60$ cm, $W=59$ cm, $H=58$ cm with single hemispherical diffractor ($h/|R|=1$) having radius (a) $|R|=5.71$ cm or (b) $|R|=10.70$ cm.}
}
\end{figure}

\begin{figure}[htb] \begin{center} \begin{tabular}{c}
\vspace{-4cm}\\
\hspace{-1.2cm}
\includegraphics[scale=0.5]{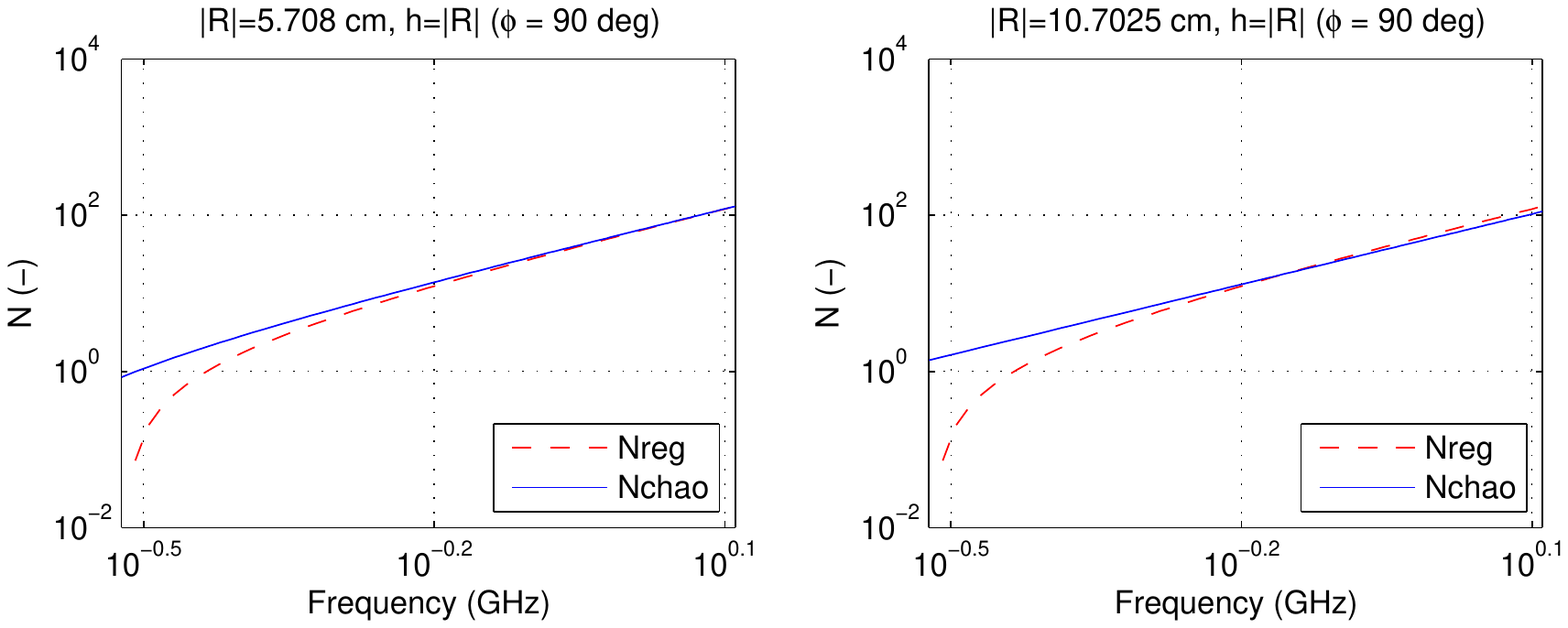}\
\vspace{-4cm}\\
\hspace{-1.2cm}
\end{tabular}
\end{center}
{
\caption{\label{fig:Nvsf15} \small
Same as Fig. \ref{fig:Nvsf1} but for $n=15$.}
}
\end{figure}

\begin{figure}[htb] \begin{center} \begin{tabular}{c}
\vspace{-4cm}\\
\hspace{-1.2cm}
\includegraphics[scale=0.5]{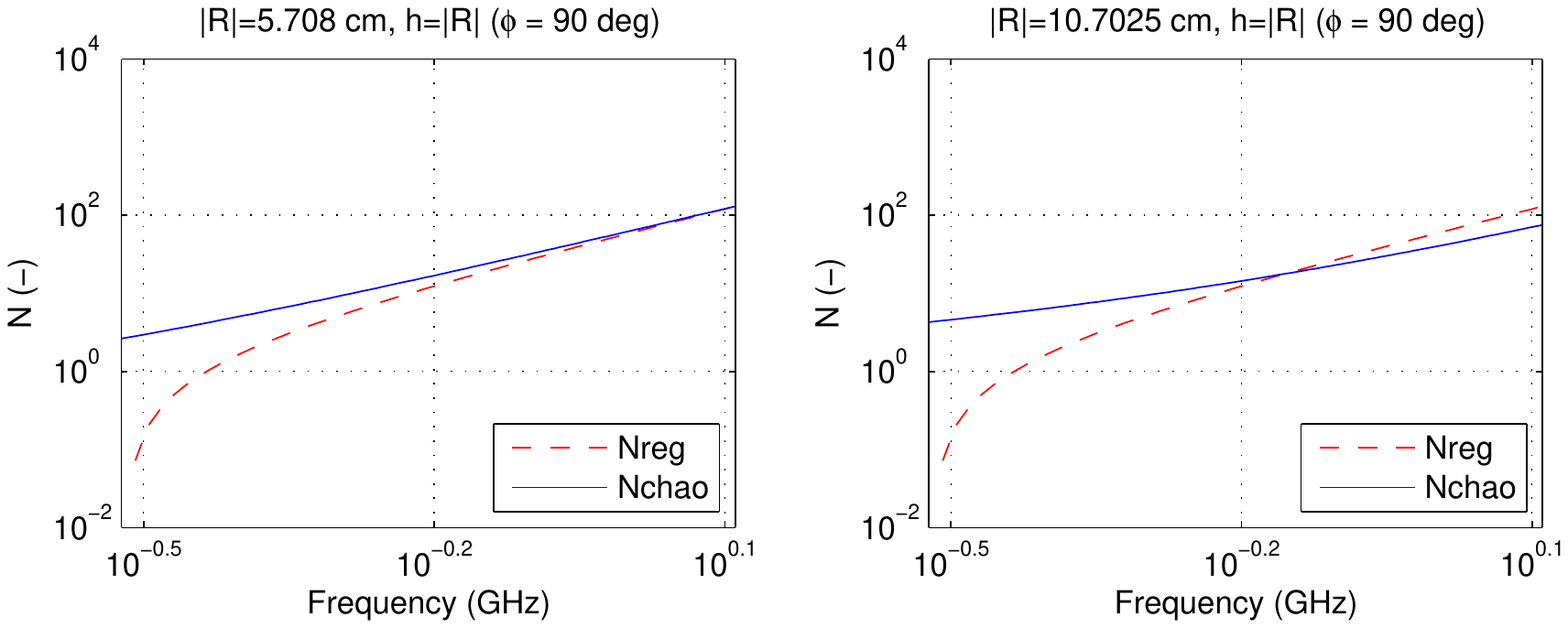}\
\vspace{-4cm}\\
\hspace{-1.2cm}
\end{tabular}
\end{center}
{
\caption{\label{fig:Nvsf45} \small
Same as Fig. \ref{fig:Nvsf1} but for $n=45$.}
}
\end{figure}

The marginal improvement by a single surface diffractor is also witnessed from the frequency characteristic $N(f)$ for fixed radius of a hemisphere, shown in Fig. \ref{fig:Nvsf1}. Even at the maximum radius for which mode increase is achieved when $f=700$ MHz, viz., $|R_{\max}|=0.2499~c/f_{\rm crit} = 10.7$ cm, the difference is still small. On the other hand, for $n=15$ (Fig. \ref{fig:Nvsf15}) and $n=45$ (Fig. \ref{fig:Nvsf45}) the increase of $N$ at low frequencies is again substantial because of aggregation of this affect.

\begin{figure}[htb] \begin{center} \begin{tabular}{c}
\vspace{-4cm}\\
\hspace{-3.2cm}
\includegraphics[scale=0.5]{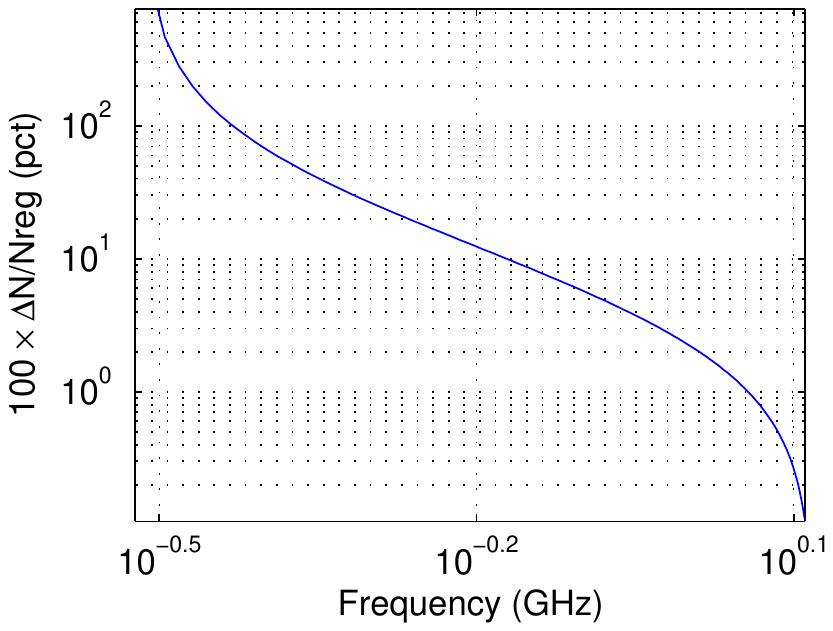}\
\vspace{-4cm}\\
\hspace{-3.2cm}
\end{tabular}
\end{center}
{
\caption{\label{fig:NvsfR5p7cm} \small
Frequency dependence of percentage increase of cumulative average mode count $N$ in lossless rectangular cavity of dimensions $L=60$ cm, $W=59$ cm, $H=58$ cm, achieved by inserting $15$ identical hemispherical surface diffractors of radius $|R|=5.71$ cm.}
}
\end{figure}

The frequency dependence of the relative increase of the mode count is shown in Fig. \ref{fig:NvsfR5p7cm} for $n=15$. The procentual change is seen to rapidly decrease with increasing frequency.


\subsection{Effect of Changing Diffractor Size $|R|$ on $N(f)$ for Fixed Diffractor Profile $h/|R|$}
\begin{figure}[htb] \begin{center} \begin{tabular}{c}
\vspace{-4.5cm}\\
\hspace{-1.2cm}
\includegraphics[scale=0.5]{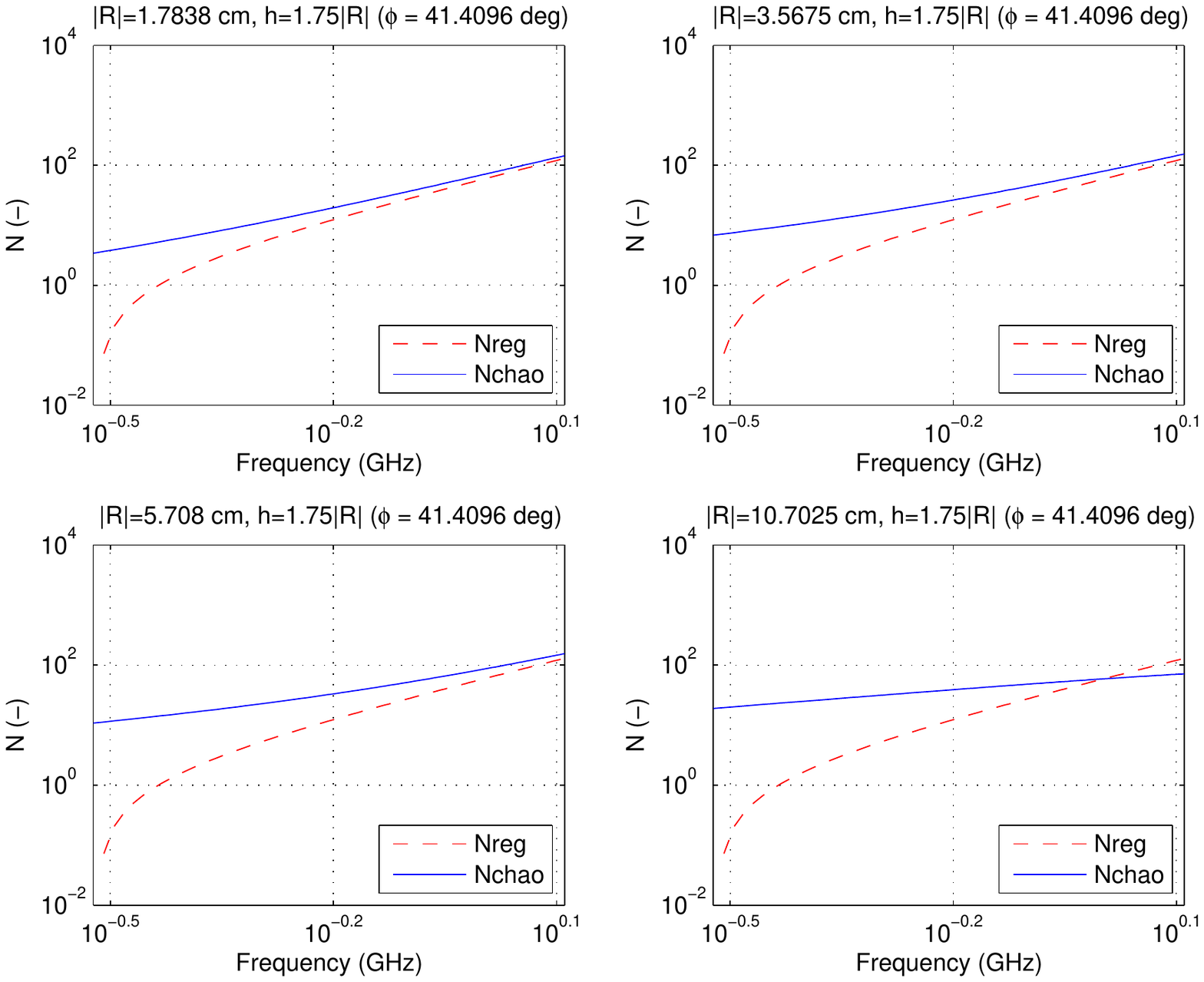}\
\vspace{-3.5cm}\\
\hspace{-1.2cm}
\end{tabular}
\end{center}
{
\caption{\label{fig:Nvsf45_hoverR1p75_paramR} \small
Frequency dependence of cumulative average mode count $N_{\rm chao}(f)$ when applying $n=45$ identical diffractors of constant relative profile height $h/|R|=1.75$, for increasing selected values of diffractor radius
(a) $|R|=1.78$ cm,
(b) $|R|=3.57$ cm,
(c) $|R|=5.71$ cm,
(d) $|R|=10.70$ cm, compared to $N_{\rm reg}(f)$ for same cavity but without diffractors. }
}
\end{figure}

\begin{figure}[htb] \begin{center} \begin{tabular}{c}
\vspace{-5.5cm}\\
\hspace{-1.2cm}
\includegraphics[scale=0.5]{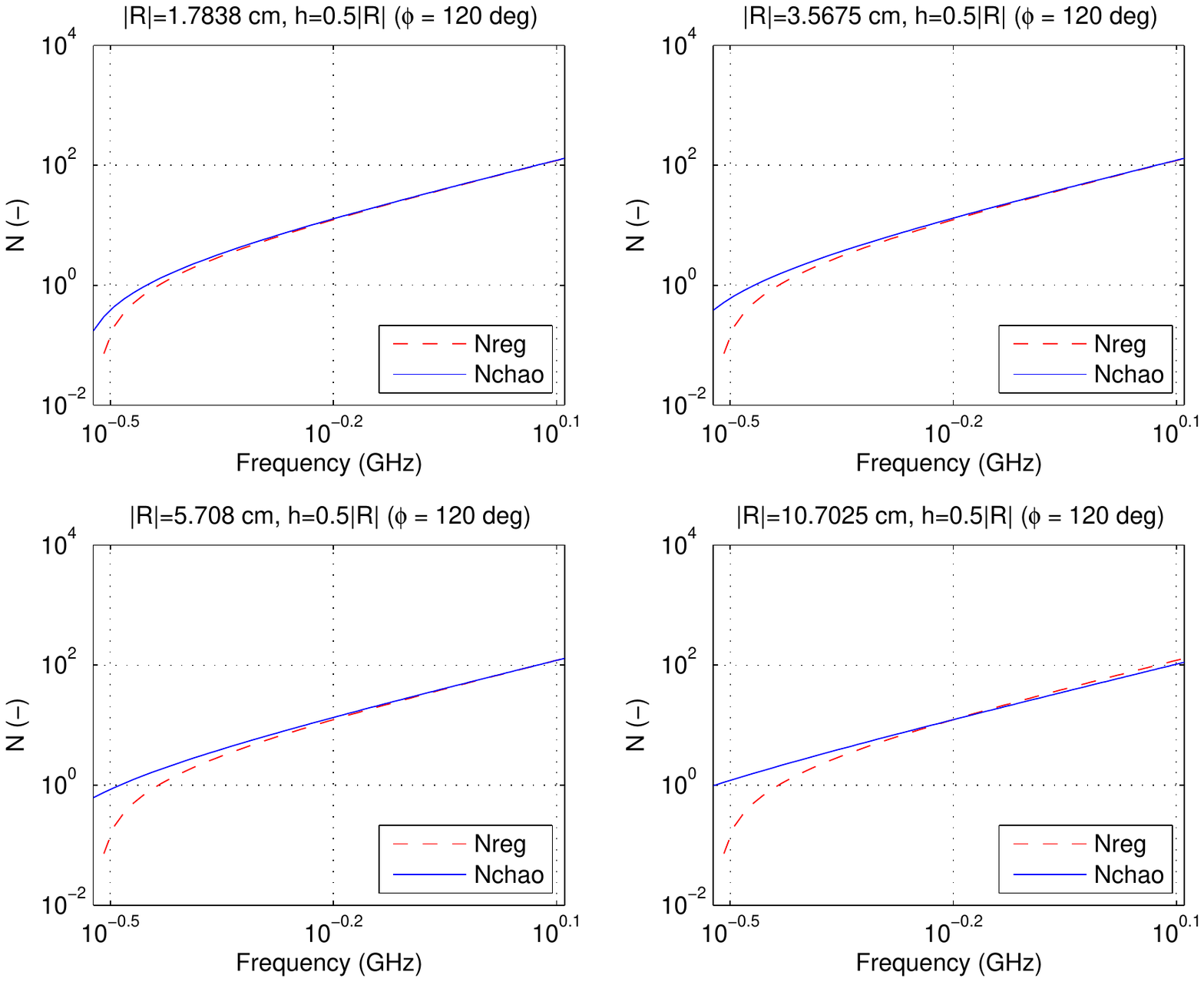}\
\vspace{-3.5cm}\\
\hspace{-1.2cm}
\end{tabular}
\end{center}
{
\caption{\label{fig:Nvsf45_hoverR0p5_paramR} \small
Same as Fig. \ref{fig:Nvsf45_hoverR1p75_paramR}, but for $h/|R|=0.5$.}
}
\end{figure}

Fig. \ref{fig:Nvsf45_hoverR1p75_paramR} shows $N(f)$ for selected values of diffractor radius, while keeping the profile height at a constant but higher value $h/|R|=1.75$ (superhemispherical diffractors). It can be seen that increasing the diffractor size (i.e., radius) has a predominant effect at lower frequencies, because diffractors that are sizeable compared to the wavelength are more efficient diffractors. The penalty is that the increase in mode density then ceases at lower frequencies.
For lower-profile (sub)hemispherical diffractors (e.g., $h/|R|=0.5$ in Fig. \ref{fig:Nvsf45_hoverR0p5_paramR}), the changes are qualitatively the same, although the improvement at low frequencies is of course more modest.

\begin{figure}[htb] \centering \begin{tabular}{c}
\vspace{-5cm}\\
\hspace{-1.2cm}
\includegraphics[scale=0.5]{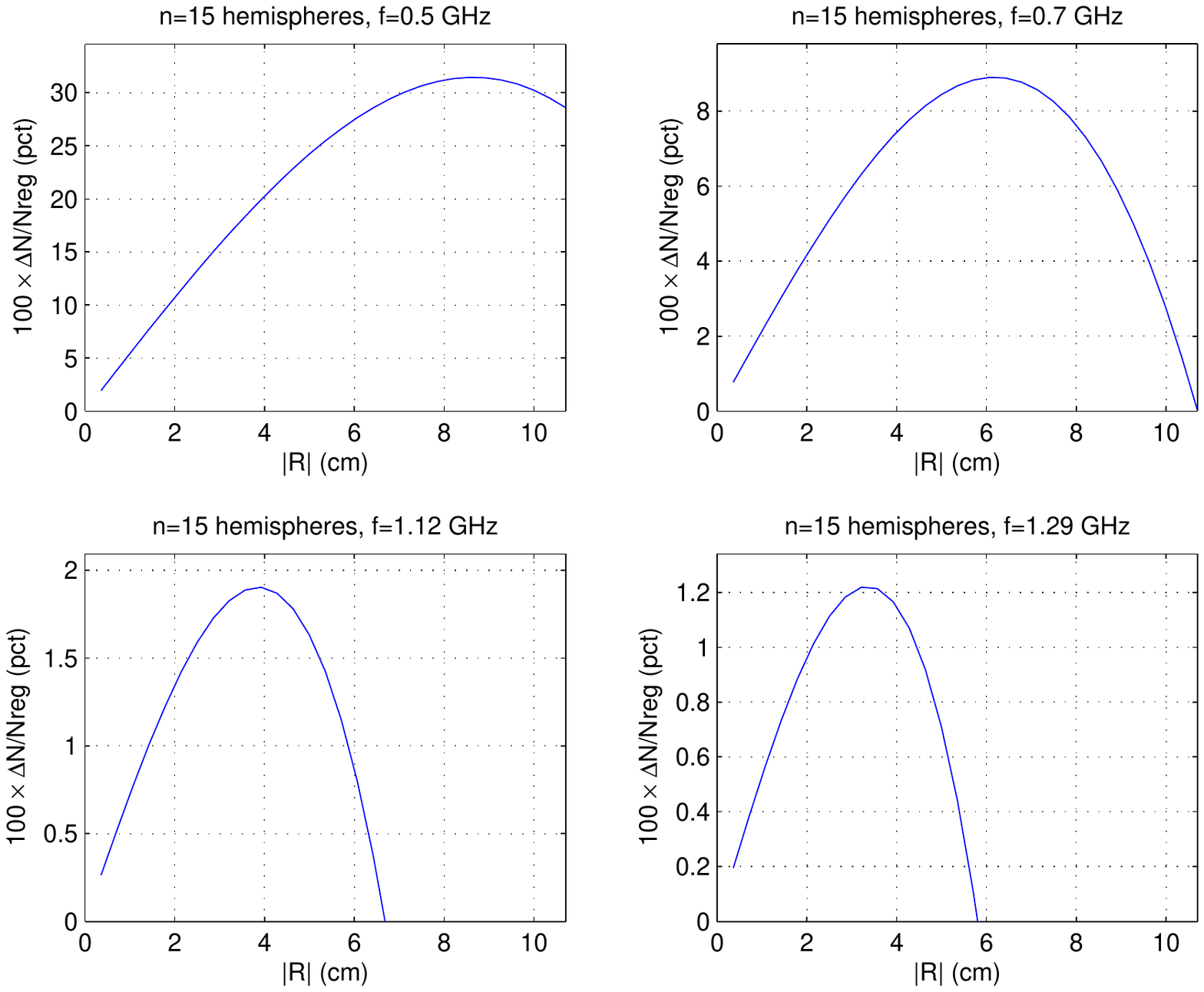}\
\vspace{-3.5cm}\\
\hspace{-1.2cm}
\end{tabular}
{
\caption{\label{fig:MED_NvsR_paramf_n15} \small
Dependence of percentage relative increase in average mode count $N$ on radius of curvature $|R|$ when inserting $n=15$ identical hemispherical diffractors ($h/|R|$), at selected frequencies. Note the different vertical scales.}
}
\end{figure}

The effect of changing $|R|$ is clearly noticeable from Fig. \ref{fig:MED_NvsR_paramf_n15} for hemispheres. Here, a frequency-dependent optimum value of $|R|$ is seen, which decreases with frequency. The effectiveness of the diffractors thus rapidly decreases for larger radii.
At the same time, at low frequencies and small radii, the increase of the relative mode count is approximately proportional to diffractor radius (see top left plot in Fig. \ref{fig:MED_NvsR_paramf_n15}), although it must be remembered that packing density (spatial density of diffractors) on a given surface decreases quadratically with increasing radius.  


\subsection{Effect of Changing Diffractor Profile Height $h/|R|$ on $N(f)$ for Fixed Diffractor Size $|R|$}
\begin{figure}[htb] \begin{center} \begin{tabular}{c}
\vspace{-4.5cm}\\
\hspace{-1.2cm}
\includegraphics[scale=0.5]{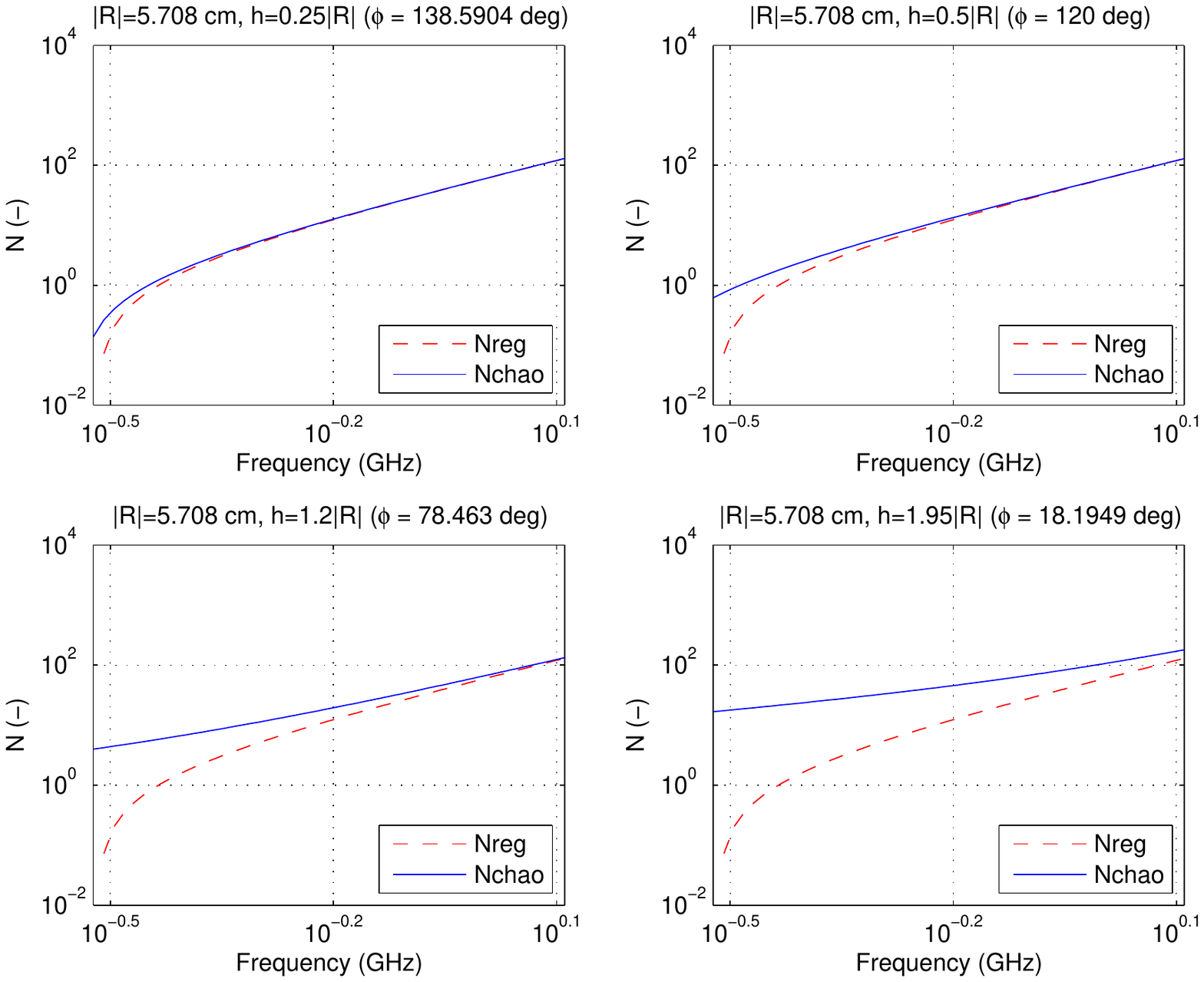}\
\vspace{-3.5cm}\\
\hspace{-1.2cm}
\end{tabular}
\end{center}
{
\caption{\label{fig:Nvsf45_R5p71cm_paramhoverR} \small
Frequency dependence of cumulative average mode count $N$ with $n=45$ for selected values of diffractor profile height, with constant radius $|R|=5.71$ cm:
(a) $h/|R|=0.25$,
(b) $h/|R|=0.5$,
(c) $h/|R|=1.2$,
(d) $h/|R|=1.95$.}
}
\end{figure}

Conversely, varying the profile height while keeping $|R|$ constant at a value $5.71$ cm has a more pronounced effect than changing $|R|$, as can be seen from Fig. \ref{fig:Nvsf45_R5p71cm_paramhoverR}. Thus, high-profile diffractors are much more efficient in increasing the LF mode density than low-profile versions.

\subsection{Effect of Rounding Edges on $N(f)$\label{sec:rounding}}
The edge term in the generalized Weyl law can produce a negative or positive contribution to $N(f)$, depending on the whether the dihedral angle $\phi$ is within or outside the interval [$\pi/5,\pi$], respectively \cite{arnaTEMC_LUF}. In particular, right angles give the (almost) worst performance, in the sense of maximally decreasing the mode density.
To alleviate this deterioration, one may consider rounding the edges such that the dihedral angle becomes $\pi$, thus removing the negative contribution of the edge term.

\begin{figure}[htb] \begin{center} \begin{tabular}{c}
\vspace{-4.5cm}\\
\hspace{-1.2cm}
\includegraphics[scale=0.5]{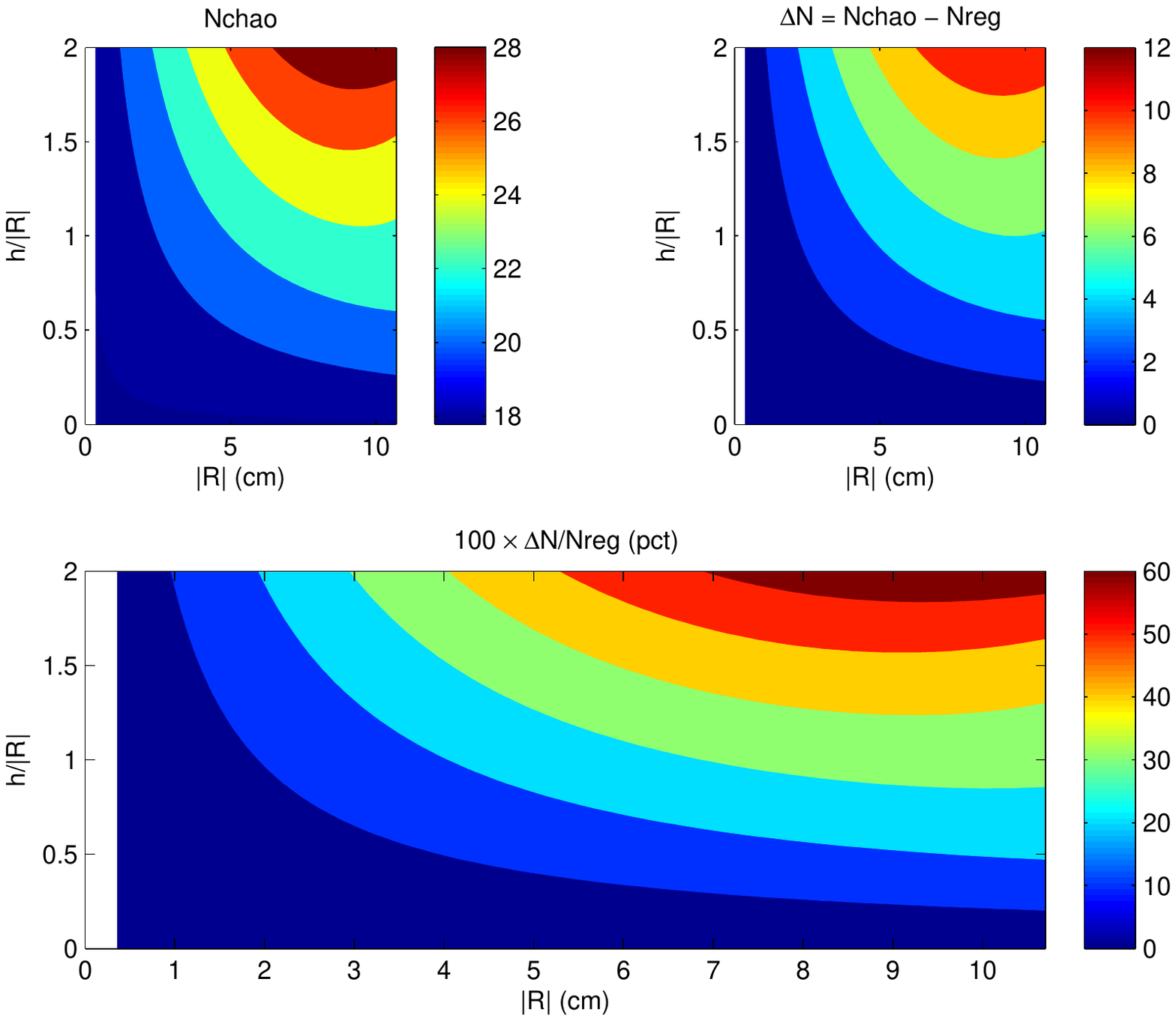}\
\vspace{-3.5cm}\\
\hspace{-1.2cm}
\end{tabular}
\end{center}
{
\caption{\label{fig:contourplot15_roundeddiff} \small
Cumulative average mode counts $N_{\rm chao}$ and $N_{\rm reg}$ and percentage relative increase of mode count at design frequency $f=700$ MHz for lossless rectangular cavity of dimensions $L=60$ cm, $W=59$ cm, $H=58$ cm furnished with $n=15$ identical convex spherical cap diffractors of radius $-|R|$ and profile height $h$, with rounded edges between diffractors and cavity surface ($\phi=180$ deg) while maintaining right angles in original cavity corners ($\phi=90$ deg).}
}
\end{figure}

\begin{figure}[htb] \begin{center} \begin{tabular}{c}
\vspace{-4.5cm}\\
\hspace{-1.2cm}
\includegraphics[scale=0.5]{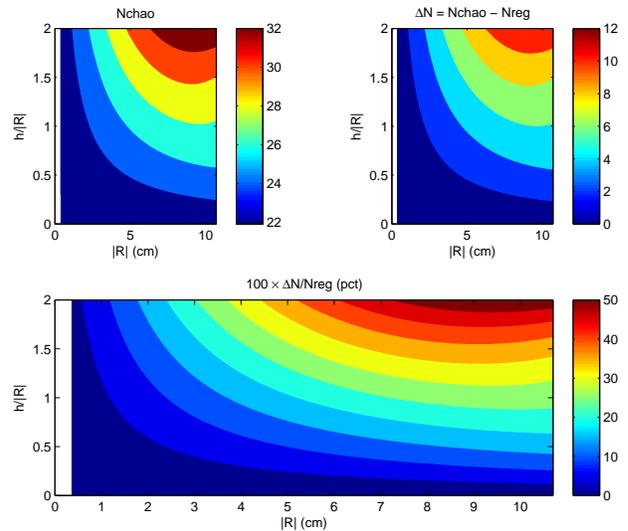}\
\vspace{-3.5cm}\\
\hspace{-1.2cm}
\end{tabular}
\end{center}
{
\caption{\label{fig:contourplot15_roundedall} \small
Same as Fig. \ref{fig:contourplot15_roundeddiff}, but with additional rounding of edges between mutually orthogonal walls of original cavity.}
}
\end{figure}

Fig. \ref{fig:contourplot15_roundeddiff} shows the cumulative mode count $N(f)$ at $f=700$ MHz for $n=15$ diffractors, now with rounded edges. These results are to be compared quantitatively to Fig. \ref{fig:contourplot15}. Note that in this calculation, the original edges of the empty cavity have not been rounded, although this would have a further beneficial effect: in the latter case, shown in Fig. \ref{fig:contourplot15_roundedall}, a levelling of the effect of diffractors on mode increase to lower profile heights and diffractor sizes is found.
The increase for near-maximum profile height is slightly reduced because of the absence of narrow dihedral angles that produce otherwise large positive edge terms \cite{arnaTEMC_LUF}.

In practical lossy cavities, the number of simultaneously excited modes at a chosen CW frequency may increase as a result of increased modal overlap occurring for increasing losses. This could make it difficult to distinguish between the effect of modal increase caused by decreased dihedral angles and the apparent increase due to losses, because of the increased charge concentration at sharper edges in the latter case. 
More generally, increased surface roughness (e.g., using aluminium foil in the practical realization of diffractors) also increases losses and modal overlap. To distinguish from an increase in mode density due to (lossless) edge contribution, an evaluation of $Q$ should provide the answer.

\clearpage

\subsection{Effect on $N(f)$ in Overmoded Regime}
\begin{figure}[htb] \begin{center} \begin{tabular}{c}
\vspace{-3.5cm}\\
\hspace{-1.2cm}
\includegraphics[scale=0.5]{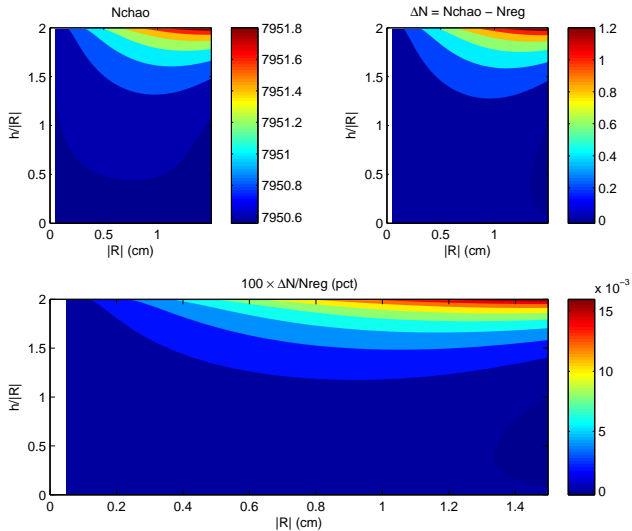}\
\vspace{-3.5cm}\\
\hspace{-1.2cm}
\end{tabular}
\end{center}
{
\caption{\label{fig:contourplot1_f5GHz} \small
Cumulative average mode count $N_{\rm chao}$ and $N_{\rm reg}$ at $f=5$ GHz for lossless rectangular cavity of dimensions $L=60$ cm, $W=59$ cm, $H=58$ cm furnished with $n=1$ single convex spherical cap diffractor of radius of curvature $-|R|$ and relative profile height $h/|R|$.}
}
\end{figure}

At relatively short wavelengths compared to the diffractor size, there is no substantial effect of introducing a curved convex wave diffractor {\em on the mode density}, as shown in Fig. \ref{fig:contourplot1_f5GHz} for $f=5$ GHz for a single diffractor ($n=1$), which is to be compared quantitatively to Fig. \ref{fig:contourplot1}: it is observed that at this higher frequency and for diffractors of radius below $1.5$ cm, the increase in mode density is now much reduced.
Of course, at these short wavelengths these diffractors assist in a different way, by increasing the local isotropy of waves passing through a single interior location as follows from semi-classical (geometric optical ray based) analysis. 


\section{Conclusions}
From the multivariable dependence of mode density on excitation frequency and various geometrical parameters, a number of features and insights emerge, for the medium sized cavity of interest:
\begin{enumerate}
\item increase of mode count or mode density in cavities with convex wave diffractors is achievable up to some critical frequency. For higher frequencies, the mode density decreases when using convex wave diffractors, because of the reduction in interior volume of the cavity;
\item the profile height $h/|R|$ is more important than the size $|R|$, everything else being equal: superhemispherical diffractors ($h/|R| > 1$) are more effective in increasing mode density than subhemispherical diffractors ($h/|R| < 1$);
\item the diffractor radius should be smaller than a quarter wavelength to achieve mode increase at frequencies above the cavity's fundamental resonance frequency (the converse case is not practically meaningful for this cavity);
\item it is difficult to achieve a substantial effect with just a single diffractor, even if large in size. However, the additive effect of arrays of smaller diffractors can be more beneficial at low to medium frequencies;
\item although smaller diffractors are individually less efficient in general, for a given surface area inside a cavity, more space is available for their placement onto (a) cavity surface(s), thus enabling a marked overall increased mode density.
Physically, the effect can be understood by the fact that the corrugation of the walls caused by the diffractors allows for a richer set of surface current patterns, compared to a cavity with flat interior walls. In turn, these surface current modes govern the interior (spatial) cavity modes on account of the EM boundary conditions. Note, however, that mode theory does not take into account the precise location of these diffractors across a wall;
\item while fluctuations of the actual low-frequency mode density around the average density dictated by the generalized Weyl law are considerable in a single static cavity \cite{bali1}, it is expected that a mechanism of ensemble averaging (mode stirring) may reduce these fluctuations on average. Such conjecture is subject to numerical and/or experimental verification.
\end{enumerate}


\appendix
\section{Appendix: Mode Density for Cavities with Hemispherical Caps\label{app:modedens}}
When adding a single convex hemispherical diffractor of radius $|R|$ to a wall of a rectangular cavity of volume $V$, this gives rise to the following changes:
\begin{itemize}
\item local curvature: $\varrho(s) = - |R|$;
\item increase in surface area: $\Delta S = (4 \pi |R|^2)/2 - \pi |R|^2 = \pi |R|^2$
\item dihedral angle along interfacial edge (rim) between diffractor and original wall surface: $\phi(\ell) = \pi/2$;
\item reduction in cavity volume: $-(4\pi |R|^3 / 3)/2 = - 2 \pi |R|^3/3$.
\end{itemize}
For $n$ such diffractors placed at random locations on any of the cavity walls, the  mode count at frequency $f$ becomes
\bea
N(f) &=& \frac{8\pi}{3 c^3} \left ( V - n \frac{2\pi |R|^3}{3} \right ) f^3 \nonumber\\
&~&
- \frac{4n}{3\pi c} \left ( \frac{2 \pi |R|^2}{-|R|} \right ) f + \frac{n}{6\pi c} \left ( -\frac{3\pi}{2}\right ) 2\pi |R| f 
\nonumber\\
&~&
+ \frac{1}{6\pi c}\left [ \left ( -\frac{3\pi}{2}\right )
\left ( 4(L+W+H)\right ) \right ] f
\\
&=&
\frac{8\pi}{3 c^3} \left ( V - n \frac{2\pi |R|^3}{3} \right ) f^3
+ \frac{16 - 3 \pi}{6c} n|R| f
\nonumber\\
&~&
-
\frac{L+W+H}{c} f \label{eq:modecountsinglediff}
\eea
where it is tacitly assumed that the diffractors are placed on the flat cavity surface without affecting (reducing) the original edge length of the parallelepiped cavity (i.e., no diffractor straddles any existing edge or corner).
The absolute change in the mode count introduced by the diffractors is thus
\bea
\Delta N &\stackrel{\Delta}{=}& N_{\rm chao} - N_{\rm reg} \nonumber\\
&=& - n \left [ \frac{16\pi^2 |R|^3}{9} \left ( \frac{f}{c} \right )^3 - \frac{(16-3\pi)|R|}{6} \left ( \frac{f}{c} \right ) \right ]~~\label{eq:Weyldiffracfinal}
\eea 
where $N_{\rm reg} \equiv N_{\rm chao}(n=0)$. 
The relative change in mode count is therefore
\bea
\frac{\Delta N}{N_{\rm reg}} &\stackrel{\Delta}{=}& \frac{N_{\rm chao}-  N_{\rm reg}}{N_{\rm reg}} \nonumber\\
&=& -n \frac{16 \pi^2 |R|^3 f^2 + \frac{3}{2} (-16 + 3 \pi) |R| c^2}{24 \pi V f^2 - 9 (L+W+H) c^2}.
\label{eq:relativechangemodedenshemi}
\eea
First, consider the case where the denominator in (\ref{eq:relativechangemodedenshemi}) is positive, i.e., at relatively short wavelengths $\lambda$ satisfying
\bea
\lambda < \sqrt{\frac{24\pi V}{9(L+W+H)}}.
\label{eq:denomcrit} 
\eea
To achieve an increase in mode density, it is then required that the numerator of the fraction in (\ref{eq:relativechangemodedenshemi}) be negative, i.e.,
\bea
16 \pi^2 |R|^3 f^2 + \frac{9\pi}{2} |R| c^2
<
24 |R| c^2
\eea
which is satisfied for
\bea
|R| < \sqrt{\frac{(24 - \frac{9\pi}{2}) c^2}{16\pi^2 f^2}} = \frac{\sqrt{24 - \frac{9\pi}{2}}}{4\pi} \lambda \simeq 0.2499 \lambda .
\label{eq:condposdenom}
\eea
Conversely, if 
\bea
\lambda > \sqrt{\frac{24\pi V}{9(L+W+H)}} 
\label{eq:denomcrit_bis} 
\eea
then the condition for increased mode density becomes 
\bea
|R| > \frac{\sqrt{24 - \frac{9\pi}{2}}}{4\pi} \lambda .
\eea

For the cavity at hand ($L\times W \times H=60\times 59\times 58$ cm$^3$), the critical value in (\ref{eq:denomcrit}) and (\ref{eq:denomcrit_bis}) is
\bea
\lambda_{\rm crit} = \sqrt{\frac{24\pi V}{9(L+W+H)}} = 0.9858~{\rm m}
\eea
corresponding to $f_{\rm crit} = 304.1$ MHz. Since this is below the first resonance of the medium sized cavity, the condition (\ref{eq:condposdenom}) applies instead.
Consequently, this poses an {\em upper\/} limit on the diffractor size as having a radius of slightly less than a quarter wavelength. (If the diffractors are of different sizes, obeying (\ref{eq:condposdenom}) is a merely sufficient condition.)
The closer $R$ approaches zero and the closer the wavelength approaches $\lambda_{\rm crit}$, the larger the positive increase $\Delta N$. For increasing frequencies and/or radii, the effect rapidly diminishes. 
In effect, when convex diffractors become too large, their negative effect on mode density brought about by reducing the cavity interior volume dominates the positive effect brought about by the increased (negative) curvature term, the latter being proportional to $|R|$.

\end{document}